\documentclass[11pt]{article}
\usepackage[margin=2cm]{geometry}
\usepackage{cite}
\usepackage{geometry}                
\usepackage{graphicx}
\usepackage{amssymb}
\usepackage{slashed}
\usepackage{amsmath,amssymb}
\usepackage{epsfig}
\usepackage{graphicx}

\usepackage{color}

\title{
\vspace*{-2cm}
\begin{flushright}
\normalsize{EFI-12-34 \\
PI-PARTPHYS-311}
~\\
\end{flushright}
\vspace*{1.5cm}
Searching for Low Mass Dark Portal at the LHC \\
\author{\textbf{Haipeng An$^a$, Ran Huo$^b$ and Lian-Tao Wang$^{b,c}$} \\
~\\
\normalsize\emph{$^a$Perimeter Institute, Waterloo, Ontario N2L 2Y5, Canada}\\
\normalsize\emph{$^b$Enrico Fermi Institute, University of Chicago, Chicago, IL 60637}\\
\normalsize\emph{$^c$Kavli Institute for Cosmological Physics, University of Chicago, Chicago, IL, 60637}\\
}}
\begin{document}
\maketitle
\vspace*{0.5cm}
\begin{abstract}

Light dark matter with mass smaller than about 10 GeV is difficult to probe from direct detection experiments. In order to have the correct thermal relic abundance, the mediator of the interaction between dark matter and the Standard Model (SM) should also be relatively light, $\sim 10^2$ GeV. If such a light mediator couples to charged leptons, it would already be strongly constrained by direct searches at colliders.  In this work, we consider the scenario of a leptophobic light $Z'$ vector boson as the mediator, and study the the prospect of searching for it at the 8 TeV Large Hadron Collider (LHC). To improve the reach in the low mass region, we perform a detailed study of the processes that the $Z'$ is produced in association with jet, photon, $W^\pm$ and $Z^0$. We show that in the region where the mass of $Z'$ is between 80 and 400 GeV, the constraint from associated production can be comparable or even stronger than the known monojet and dijet constraints. Searches in these channels can be complementary to the monojet search, in particular if the $Z'$ couplings to quarks ($g_{Z'}$) and dark matter ($g_D$) are different. For $g_D < g_{Z'}$, we show that there is a larger region of parameter space which has correct thermal relic abundance and a light $Z'$, $M_{Z'} \sim 100 $ GeV. This region, which cannot be covered by the mono-jet search, can be covered by the resonance searches described in this paper.
\end{abstract}
\thispagestyle{empty}
\newpage

\section{Introduction}

Dark Matter (DM) consists of 24\% of the energy density of our universe. However, the nature of it is one of the outstanding mysteries. The most popular class of candidates for DM are the stable Weakly Interacting Massive Particles (WIMPs), and the annihilation of WIMPs in the early universe determined the observed relic abundance of DM. This scenario provides a promising approach to detect the WIMPs directly~\cite{Goodman:1984dc}, with many recent results~\cite{Aalseth:2011wp,Akerib:2010pv,Ahmed:2010wy,Aprile:2010um,Angloher:2011uu,Bernabei:2008yi,Aprile:2011hi,Felizardo:2010mi,Aprile:2012nq}. The direct detection experiments are more sensitive for WIMPs with masses heavier than tens of GeV. Assuming the WIMP-nucleon interaction is spin independent, the limit can be as strong as $ \sigma_{\rm WIMP-nucleon} < 10^{-43}$ to $10^{-45}$ cm$^2$. For light WIMPs, the energy transfer during a WIMP-nucleus scattering is proportional to the WIMP mass. If the WIMP is sufficiently light, the energy transfer is too small to be detected, resulting in a much weaker limit. For example, the upper limit is weaker than $10^{-39}$ cm$^2$ for $M_{\rm WIMP} \sim 5$ GeV.

Recently, the low mass dark matter scenario has received more attention. For example, there is an intriguing connection between the WIMP number density and baryon number density in this scenario~\cite{Nussinov:1985xr,Kaplan:2009ag,Farrar:2005zd,Kitano:2004sv,Agashe:2004ci,An:2009vq,Shelton:2010ta,Davoudiasl:2010am,Bell:2011tn,Gu:2010ft,Blennow:2010qp,
Dutta:2010va,Kang:2011wb,Cheung:2011if,MarchRussell:2011fi,Frandsen:2011kt,Ibe:2011hq,Kamada:2012ht,Feng:2012jn}, which is motivated by the fact that the energy densities of dark matter and baryons are of the same order of magnitude. High energy colliders can play a significant role in the search for light WIMPs. The simplest approach is to assume that the interaction between WIMPs and Standard Model (SM) particles can be written as effective contact operators~\cite{Goodman:2010yf,Bai:2010hh,Goodman:2010ku,Fortin:2011hv,Rajaraman:2011wf,Shoemaker:2011vi,Graesser:2011vj,Friedland:2011za}, with the mediators between the WIMP and the SM sector 
integrated out. The simplest observable signal would be WIMP pair production associated with a jet, which is often referred as the monojet + missing transverse energy (MET) search channel.

On the other hand, it is a generic possibility that the mediator can be light. In this case, it cannot be integrated out while considering scattering process at the LHC. Such an effect has been studied in Refs.~\cite{Bai:2010hh,Goodman:2010ku,Fortin:2011hv,Rajaraman:2011wf,Shoemaker:2011vi,Graesser:2011vj,Friedland:2011za,An:2012va}. A light mediator is particularly motivated by reproducing the observed thermal relic abundance, which requires a WIMP annihilation cross section is of order picobarn. If WIMP annihilation mainly proceeds through s-wave , the cross section can be written as
\begin{equation}\label{sigmav}
\langle\sigma v\rangle \approx \frac{ g^2_{Z'} g_D^2 N_C N_f M_D^2}{2\pi M_{Z'}^4}=3\times10^{-26}\left(\frac{g_{Z'}}{0.27}\right)^2\left(\frac{g_D}{0.27}\right)^2\left(\frac{M_D}{5\mbox{GeV}}\right)^2\left(\frac{100\mbox{GeV}}{M_{Z'}}\right)^4\thinspace\mbox{cm}^3\mbox{s}^{-1},
\end{equation}
where we have assumed that the interaction between WIMP and the SM quarks is mediated by a vector boson $Z'$. This is the so called dark portal. The relevant Lagrangian can be written as
\begin{equation}
\mathcal{L}\ni g_{Z'}\bar{q}\gamma^\mu q Z_\mu' + g_D\bar{\chi}\gamma^\mu\chi Z_\mu',
\end{equation}
where we have assumed that the dark matter particle, denoted by $\chi$, is a Dirac fermion.
$g_{Z'}$ and $g_D$ are the coupling of the mediator to quarks and to WIMP, respectively. $N_f$ is the number of SM \emph{flavors} that are kinematically available. For such a low mass resonance its coupling to leptons must be strongly suppressed, otherwise it would be ruled out by LEP. That's the reason that in our model the $Z'$ couples only to quarks in SM (leptophobic).

Such a light mediator with $M_{Z'} \sim \mathcal{O}(100)$ GeV  is within the reach of the Large Hadron Collider (LHC). The most straightforward way to search for a leptophobic $Z'$ is a direct search for dijet resonance, a ``bump hunt'', which is summarized in Ref.~\cite{An:2012va}. Although there is some model dependence, it turns out that resonance searches for the vector mediator with masses between 250 GeV and 4 TeV at the colliders and mapping to DM direct detection constraint can provide stronger constraints than the monojet searches. On the other hand, the constraints become much weaker for low mass mediator. Since the jets from lighter $Z'$ decay would be softer, the signal of a light $Z'$ suffers from very low trigger efficiencies.

In this work, we study  a class of alternative channels in which the $Z'$ boson is produced in association with a hard jet,  a hard photon, or a massive gauge boson in the SM. We also consider the case of $Z'$ pair production. The rates of the SM background for the associated productions can be significantly lower. Therefore, we can lower the trigger threshold and enhance the signal efficiency. In the following we will demonstrate that this is indeed a promising way of finding or constraining the leptophobic $Z'$ mediator, and eventually the DM particle.

We present a detailed description of the Monte Carlo simulation in Sec.~2. In Sec.~3, we scan the parameter space and show the reach for each channel and their combination. In Sec.~4, we relate the $Z'$ search to  the collider search of dark matter.  We summarize the results in the Sec.~5.

\section{Simulation and Event Selection}

\subsection{Signal and Background Simulation}

The signal and background are both generated by \texttt{MadGraph/MadEvents 5.1.3.9} \cite{Alwall:2011uj}. We used the $k_T$ MLM Matrix Element-Patron Shower (ME-PS) matching algorithm. The decay width of $Z'$ is calculated by \texttt{BRIDGE 2.24} \cite{Meade:2007js}. The following decay and hadronization procedure are performed by \texttt{pythia-pgs 2.1.8}. The jets are constructed by \texttt{fastjet 3.0.0} \cite{Cacciari:2011ma} with $R=0.6$ anti-$k_T$ jet algorithm, and smearing is performed for jets, leptons, photons with the CMS-like energy resolution \cite{Ball:2007zza}
\begin{eqnarray}\label{smear}
\frac{\delta E_J}{E_J}&=&\frac{0.84}{\sqrt{E_J/\mbox{GeV}}}\oplus0.07,\\
\frac{\delta E_\ell}{E_\ell}&=&\frac{0.028}{\sqrt{E_\ell/\mbox{GeV}}}\oplus0.003\oplus\frac{0.125}{E_J/\mbox{GeV}},\\
\frac{\delta E_\gamma}{E_\gamma}&=&\frac{0.01}{\sqrt{E_\gamma/\mbox{GeV}}}\oplus0.007.
\end{eqnarray}
Throughout this paper we assume a 15 fb$^{-1}$ total integrated luminosity at centre-of-mass energy of 8 TeV.

The next-to-leading order correction K-factor $K=\sigma_{NLO}/\sigma_{LO}$ will not significantly change our results. In particular, it is calculated by \texttt{MCFM 6.2} \cite{Campbell:MCFM} to be roughly one, both for the $Z'\thinspace+W^\pm$ channel and the $Z'\thinspace+Z^0$ channel. For the $Z'\thinspace+$ jet and the $Z'\thinspace+\thinspace\gamma$ channels we do not preform a calculation, but \cite{Bern:2011ep} suggests the K factor for pure QCD 3 jets should be about 1.3 or so, and in \cite{Bern:2011pa} the K factor for $\gamma+$ 2 jets is about 1.2 or so.
All the K factors are close to one, so we expect counting the higher order corrections will not significantly change our reach.

We approximate $Z'$ as a Breit-Wigner resonance. The decay width of $Z'$  can be written as
\begin{equation}\label{width_ana}
\Gamma\simeq \frac{g_{Z'}^2}{12\pi}M_{Z'}(N_f N_C + r^2),
\end{equation}
where $N_f=5$ or 6 if $M_{Z'}$ is above the ditop threshold, $N_C=3$ is the number of colors in QCD, and $r\equiv g_D/g_{Z'}$.
As we will see in the following sections, the narrow width condition $\Gamma_{Z'}\ll M_{Z'}$ is always satisfied in the parameter region we are interested in. Therefore, the Breit-Wigner approximation is valid in our case. We will not consider the interference effect of the $Z'$ signal with the  SM background. Such an effect is only important within the $Z'$ width. So in our narrow width $Z'$ scenario the effect is small, and it will be completely washed out due to the jet energy resolution. We verified that this is indeed the case.

\subsection{Event Selection}

The most important class of event selection cuts for our signal are the jet energy thresholds and acceptance cuts. We begin with the $Z'$ + jet channel. Due to the large QCD background, we have to adopt relatively high jet energy threshold. With our choice, the leading order SM background cross section corresponds to an event rate of a few Hz or so, with the current peak instantaneous luminosity. Since the energy of the jets from $Z'$ decay is closely correlated with the $Z'$ mass, we use two complementary sets of $p_{TJ}$ selection cuts, so that in combination they give us good signal efficiency for a large range of $M_{Z'}$. For light $Z'$ ($M_{Z'}<350$ GeV) we require
\begin{equation}\label{QCDlight}
p_{TJ1}>350\enspace\mbox{GeV}\qquad\qquad p_{TJ2},p_{TJ3}>70\enspace\mbox{GeV}.
\end{equation}
In this case, the second and third hardest jets are dominantly coming from $Z'$ decay. We choose to use a relatively low threshold for them to enhance signal efficiency. At the same time, we have to require the hardest jet to be very energetic to suppress the rate of QCD background.
For heavier $Z'$ ($M_{Z'} > 350$ GeV), the decay products of $Z'$ give two hard jets that are most likely to be the two leading jets. The two hardest jets, if coming from $Z'$ decay, are typically close in $p_T$. Therefore, equal thresholds for the first two leading jets are desirable. We also require the jet produced in association with $Z'$ to be energetic to suppress the QCD background. In particular, we require
\begin{equation}\label{QCDheavy}
p_{TJ}>160\enspace\mbox{GeV for three leading jets}.
\end{equation}

In the other channels, $Z' + \gamma$ and $Z' + W^\pm/Z^0$, the signal also contains at least two jets coming from $Z'$ decay. At the same time, the additional hard object in the event, such as a hard photon or hard leptons from $W^\pm/Z^0$ decay, can be used to efficiently  trigger on this class of events. Therefore, we can afford to use lower thresholds for jets
\begin{equation}\label{PTJCut}
p_{TJ}>50\enspace\mbox{GeV for two leading jets}.
\end{equation}

We will focus on central jets with a good energy resolution.  For all the channels considered in this paper, we impose an acceptance cut
\begin{equation}\label{EtaJCut}
|\eta_{J}|<2.5.
\end{equation}
We apply this cut not only to jets but also to $\gamma$s and charged leptons.

We also note that the two jets from $Z'$ decay tend to be central. At the same time, the background QCD jets are more forward with larger rapidity gaps.
Therefore, we impose an $\eta$ separation cut
\begin{eqnarray}\label{EtaJSeparation}
|\Delta\eta_{JJ}|<1.7,
\end{eqnarray}
for the two jets which are used to reconstruct the $M_{Z'}$.

We have also considered different jet energy thresholds in both light and heavier $Z'$ cases. The results are tabulated in Table~\ref{TrigVar_1} and Table~\ref{TrigVar_2} in Appendix~\ref{sec:jet_trig}. While we use the relatively conservative choices of the jet energy threshold, lower them can certainly increase the reach.

\begin{table}\label{InvMass}
  \begin{center}
    \begin{tabular}{|c|c|c|c|c|c|c|c|}
      \hline
      $Z'$ Mass (GeV) & 60 & 80 & 100 & 120 & 150 & 250 & 350 \\
      \hline
      $M_{JJ}$ (GeV) & 50-70 & 60-90 & 80-110 & 100-130 & 120-160 & 210-270 & 290-370 \\
      \hline
      $Z'$ Mass (GeV) & 450 & 550 & 650 & 750 & 850 & 950 & \\
      \hline
      $M_{JJ}$ (GeV) & 370-480 & 450-590 & 530-700 & 610-800 & 700-900 & 800-1000 & \\
      \hline
    \end{tabular}
    \caption{Invariant mass window for each hypothetical $Z^\prime$ mass. We optimize this with signal Monte Carlo, where the boundaries correspond to the bins with half of the peak bin height. 
      }
  \end{center}
\end{table}

To search for the $Z'$ resonance, we have to identify two jets as candidates for $Z'$ decay products. In the $Z' + $jet channel with  $M_{Z'} < 350$ GeV, we choose the second and third hardest jet. For $M_{Z'} > 350 $ GeV, we choose the two leading jet instead. In $Z' + \gamma$ and $Z' + W^\pm/Z^0$ channels, we also choose the two leading jets. After identifying the two candidate jets, we require their invariant mass to be within the mass window around the target $Z'$ mass.  The width of the window has two origins, one is the natural line width or the physical $Z'$ total decay width, and the other is due to the detector finite energy resolution. In our case,  the latter one  always dominates.
We optimized the mass window for a set of $Z'$ masses based on Monte Carlo simulation. The complete list of the invariant mass windows used in our analysis is Table~\ref{InvMass}.

We also impose selection cuts on the other (non-jet) hard objects in the signal. For the $Z^\prime$ + $\gamma$ channel, we require
\begin{equation}
p_{T\gamma}>50\enspace\mbox{GeV}.
\end{equation}
For the $Z^\prime$ + $W^\pm$ channel, we focus on the leptonic channel. We require
\begin{eqnarray}
p_{T\ell}>25\enspace\mbox{GeV},\\
p\hspace{-0.4em}\slash_{T}>25\enspace\mbox{GeV}.
\end{eqnarray}
For the $Z^\prime$ + $Z^0$ channel, we consider two different $Z^0$ decay modes. The first channel is 
the charged leptonic $Z^0$ decay mode. In this case, there are two opposite sign electrons or muons which correctly reconstruct a $Z^0$. Specifically, we require
\begin{eqnarray}
p_{T\ell}>25\enspace\mbox{GeV for two opposite sign leptons},\\
85\enspace\mbox{GeV}<M_{\ell\ell}<97\enspace\mbox{GeV}.
\end{eqnarray}
The second channel is 
the invisible decay $Z^0>\nu\tilde{\nu}$. We require
\begin{equation}
\slashed{p}_{T}>60\enspace\mbox{GeV},
\end{equation}
and 
veto any charged leptons. 
They should be replaced by the practical triggering conditions at the LHC.

In our study, we have not considered $Z' \to b \bar b$ as a possible decay channel, mainly for simplicity. The results are already quite encouraging without $b\bar b$ decay channel,
and including it will certainly enhance the discovery reach. Extending the analysis in this paper to this case is straightforward, after properly taking into account the b-tagging. For heavier $Z'$, decaying into $t \bar t$ would also give another signal channel.
At the same time, identifying the top requires different strategies, depending on $M_{Z'}$ \cite{Altheimer:2012mn}. %

\section{Reach of Different Channels}

In general, the couplings of $Z'$ to the left and right-handed quarks can be different. In the $Z'\thinspace+$ jet and $Z'\thinspace+\thinspace\gamma$ channels, since QCD and QED are vector-like, only the combination $g=(g_{Z'L}^2+g_{Z'R}^2)^{1/2}$ is relevant, where $g_{Z'L}$ and $g_{Z'R}$ are the couplings of $Z'$ to the left and right-handed quarks. Only the left handed coupling is relevant  for the $Z'\thinspace+\thinspace W^\pm$ channel. The most complicated channel is $Z'\thinspace+\thinspace Z^0$, in which both the $g_{Z'L}$ and $g_{Z'R}$ are relevant. Further complication comes with two kinds of decay channels under consideration. However, we will see in the following that the reach is mainly due to the non-chiral $Z'\thinspace+$ jet channel. The chiral $Z'\thinspace+\thinspace W^\pm$ contribution is nearly the same as the non-chiral $Z'\thinspace+\thinspace\gamma$ contribution.  $Z'\thinspace+\thinspace Z^0$ has a very small contribution. Therefore, we expect the ratio  $g_{Z'L}/g_{Z'R}$ will not play a significant role, and we will only show the results of $g_{Z'L}=g_{Z'R}\equiv g_{Z'}$.

\begin{table}[h!]
\centering
\begin{tabular}{|c|c|c|c|c|c|}
\hline
& QCD 3 jet& dijet$+ \gamma$ & dijet$+W^\pm$ & dijet+$(Z^0 \to \ell^+\ell^-)$ & dijet$+(Z^0 \to \nu\tilde{\nu})$ \\
\hline
$p_{TJ}$, $\eta$ cuts & $\hspace{-0.4em}\begin{array}{l}487~\mbox{pb} \\ 412~\mbox{pb} \end{array}$ & 327~pb & 160~pb & 7.6~pb  & 20~pb \\
\hline
$\Delta\eta$ cut & $\hspace{-0.4em}\begin{array}{l}370~\mbox{pb} \\ 284~\mbox{pb} \end{array}$ & 226~pb  & 109~pb  & 4.2~pb  & 14~pb \\
\hline
$M_{JJ}\in[120-160]$ GeV & 42~pb & 49~pb & 13~pb & 0.95~pb & 2.5~pb \\
$M_{JJ}\in[290-370]$ GeV & 67~pb/11~pb & 18~pb & 4.0~pb & 0.32~pb & 1.3~pb \\
$M_{JJ}\in[800-1000]$ GeV & 25~pb & 0.76~pb & 0.21~pb & 0.04~pb & 0.11~pb \\
\hline
\end{tabular}
\caption{\label{tab:channel} Leading order standard Model background rate for each of the channels, after various selection cuts. In the first row the basic $p_T$ thresholds and $\eta<2.5$ cut for \emph{all particles} in an event are implemented. Then in the second row we add $\Delta\eta_{12}$ cuts for two leading jets, as well as the invariant mass window for $e^+e^-$ or $\mu^+\mu^-$ in the dijet+ $(Z^0 \to \ell^+\ell^-)$ channel to reconstruct $Z^0$. The event selection cuts are designed in particular to make the corresponding rates here no more than a few Hz for each channel. Lastly invariant mass windows are applied for three hypothetical $Z'$ masses. In the QCD 3 jet channel, we list the two cross sections, which correspond separately to the 350-70-70 $p_{TJ}$ triggering (the former) and the 160-160-160 $p_{TJ}$ triggering (the latter).}
\end{table}
\medskip
\begin{table}[h!]
\centering
\begin{tabular}{|c|c|c|c|c|c|}
\hline
& $Z'+$ jet & $Z'+ \gamma$ & $Z'+W^\pm$ & $Z'+(Z^0 \to \ell^+\ell^-)$ & $Z'+(Z^0 \to \nu\tilde{\nu})$ \\
\hline
$M_{Z'}=150$ GeV & 0.80~pb & 0.76~pb & 0.45~pb & 0.031~pb & 0.054~pb \\
$M_{Z'}=350$ GeV & 0.72~pb/0.15~pb & 0.30~pb & 0.14~pb & 0.0096~pb & 0.021~pb \\
$M_{Z'}=950$ GeV & 0.20~pb & 0.013~pb & 0.006~pb & 0.0004~pb & 0.0011~pb \\
\hline
\end{tabular}
\caption{\label{channel_sig} Leading order $Z'$ signal rate for each of the channels. All selection cuts in Table 2 (including the mass window) are applied.
Here we are assuming $g_L=g_R=0.35$. Again in the $Z'\thinspace+$jet channel we list the two cross sections, which correspond separately to the 350-70-70 $p_{TJ}$ threshold (the former) and the 160-160-160 $p_{TJ}$ threshold (the latter).}
\end{table}

\begin{figure}
\centering
\includegraphics[height=3in]{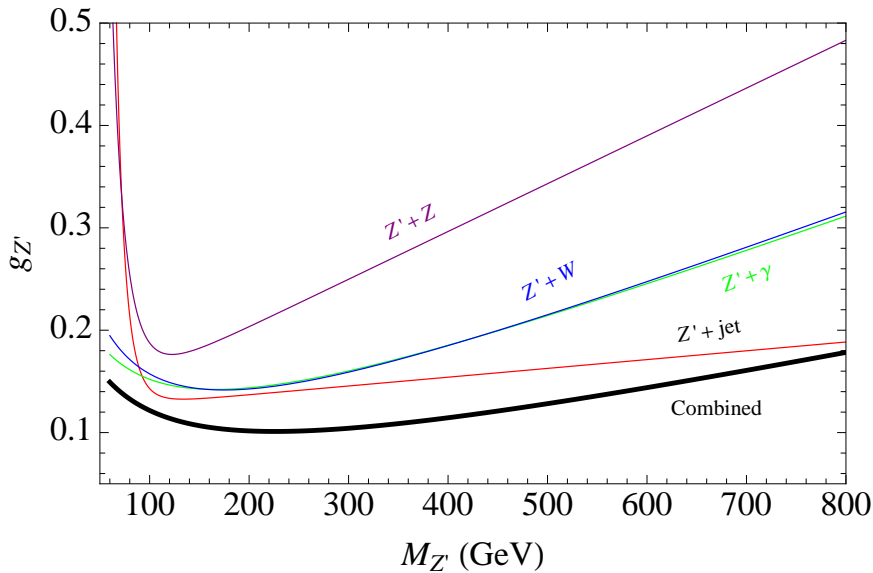}
\caption{Reach plots for $S/\sqrt{B}$ for the $Z'$ + jet (red), the $Z'\thinspace+\thinspace\gamma$ (green), the $Z'\thinspace+\thinspace W^\pm$ (blue) and the $Z'\thinspace+\thinspace Z^0$ (purple) channels, for 15 fb$^{-1}$ integrated luminosity and $S/\sqrt{B}=2$ which corresponds to 95\% confidence level. All the channel is assuming $g_{Z'L}=g_{Z'R}=g_{Z'}$, and in the $Z'\thinspace+\thinspace Z^0$ channel the reach from two charged leptons decay product and from two neutrinos are combined. Eventually their combination for exclusion is shown as the black thick line.}\label{fig:Z4channels}
\end{figure}

As illustration, a list of cross sections for various background channels and signal channels with $g_{Z'}=0.35$ are shown separately in Table~\ref{tab:channel} and Table~\ref{channel_sig}.
The expected $2\sigma$ constraints from different channels are shown in Fig.~\ref{fig:Z4channels}, where the red, green, blue, purple and black curves correspond to the upper limits from the $Z'+{\rm jet}$, $Z'+\gamma$, $Z'+W^{\pm}$, $Z'+Z^0$ channels and the combined constraint of the four channels, respectively. Here we have assumed here $g_D = 0$. In realistic models considered later, the decay branching ratio to jets will induce a suppression to the signal rate in each channel, however, the suppression is small at least for $g_D\sim g_{Z'}$\footnote{Note that due to the large number of degrees of freedom in the quark sector, and our assumption that there is only one species of dark matter, $g_D = g_{Z'} $ only leads to a small invisible width, BR$_{\rm inv} \sim 1/16$.}. A detailed discussion is in Sec.~\ref{sec:darkportal}. For $M_{Z'}\gtrsim1000$ GeV the background QCD dijet rate is low enough that the more efficient direct dijet resonance search provides a better limit~\cite{Aad:2011fq}, so we only consider the cases with $M_{Z'} < 1000$ GeV here.

From Fig.~\ref{fig:Z4channels}, we can see that the constraint from $Z'+$jet channel is stronger than other channels, especially in the large $M_{Z'}$ region. The dominant reason is the QCD coupling is much larger than the other relevant gauge couplings. For the $Z'+\gamma$ channel, both the signal and the background are suppressed by the fine structure constant of electromagnetic interaction, $\alpha_{\rm em}$. Whereas in the case of $Z'$+jet channel, it is replaced by $\alpha_S$, which is much larger. As a result, both the signal and the background increase by roughly the same amount in comparison with the $Z'+\gamma$ channel, and therefore the $S/\sqrt{B}$ for the $Z'$+jet channel is larger than that for the $Z'+\gamma$ channel. This is exactly what happens in the large $M_{Z'}$ in Fig.~\ref{fig:Z4channels}. In the small $M_{Z'}$ region, as shown in Tables~\ref{tab:channel} and \ref{channel_sig}, the cuts make the cross sections of $Z'+$jet channel and $Z'+\gamma$ channels comparable in both signal and background processes. Therefore, as shown in Fig.~\ref{fig:Z4channels}, the constraints on $g_{Z'}$ is comparable in the region $M_{Z'}\sim150$ GeV.

In the $Z'+$jet channel, in the small $M_{Z'}$ region where $M_{Z'} \lesssim 100$ GeV, to balance the transverse momentum of the hardest jet, the $Z'$ resonance is highly boosted. Therefore, the two jets from the decay of the $Z'$ resonance are  close to each other, and cannot be distinguished from a single fat jet. Due to the same reason, the small $M_{23}$ region in the background is also removed, where $M_{23}$ is the invariant mass of the second and third hardest jets. As a result, the events of both the signal and the background are cut off by the requirements in Eq.~(\ref{QCDlight}) and the statistics becomes poor, and therefore the limit on $g_{Z'}$ from $Z'+$jet channel weakens significantly in the region where $M_{Z'}\lesssim 80$ GeV. A quantitative analysis which shows the scale of this effect can be found in Appendix~\ref{appendixB}.  For example, for $M_Z' \sim 80$ GeV which is discussed in Appendix~\ref{appendixB}, if we use the $R=0.1$ anti-$k_T$ jet algorithm rather than the default $R=0.6$ one in a hadron level analysis, we will get about 20 times more events with a dijet invariant mass in the window of $60-90$ GeV. Jet substructure techniques should be able to help in this regime, and we leave the details in this direction for a future study.
In contrast, for the $Z'+\gamma$ and $Z'+W$ channels, there is no such kinematic configuration. Therefore, in the small $M_{Z'}$ region, the constraint is mainly from these two channels.

In the $Z'\thinspace+\thinspace\gamma$ channel, it is possible for jets to fake photons. For example, there can be a hard jet which accidentally becomes a $\pi^0$ and passes the neutral electromagnetic trigger to mimic a photon. The jet fake gamma rate is tiny (less than 0.3\%), but given the QCD jets are much more abundant, the fake trigger is still considerable. Taking this into consideration, the background will increase by about 20\%.

The $Z'\thinspace+\thinspace W^\pm$ channel has comparable $S/\sqrt{B}$ ratio with the $Z'\thinspace+\thinspace\gamma$ channel over all the $Z'$ mass region, which is just a coincidence of various physical reasons contributing in different directions. For example, the weak coupling is larger than the electromagnetic coupling, the $W^\pm$ is massive so that the production is relatively suppressed.  $W^\pm \to \ell^{\pm} $ has further suppression from the leptonic decay branching ratio. At the same time, the $\gamma$s are concentrated in the collinear region and hard to pass the $p_{T\gamma}>50$ GeV selection cut, while the acceptance of a $W^\pm \to \ell^{\pm}$ is higher.

For the $Z'\thinspace+\thinspace Z^0$ channel, we have to combine its two decay channels. The neutrino channel has a larger branching ratio ($\sim$20\%) than the chagerd lepton channel ($\sim$6.7\%), and  the former has a slightly better $S/\sqrt{B}$. However, even the combined signal significance is much smaller than other channels, so this channel is less interesting.

We have also checked the $Z'\thinspace+\thinspace Z'$ pair production channel. The $S/\sqrt{B}$ ratio is always much less than 1 in the region of mass and coupling we focus on. One reason is that the couplings are all relatively small, which leads to small production cross section. At the same time,  the signal is in  a pure 4-jet final state,  which is overwhelmed by the dominant QCD background. For a similar background rate with the 3-jet case we have to use nearly the same jet $p_{TJ}$ thresholds, but the signal cross section is further suppressed by a small factor of $g_{Z'}^2$. Therefore, the reach in this channel would be much weaker,  and we will not provide the full analysis here.

\begin{figure}[h!]
\centering
\includegraphics[height=3in]{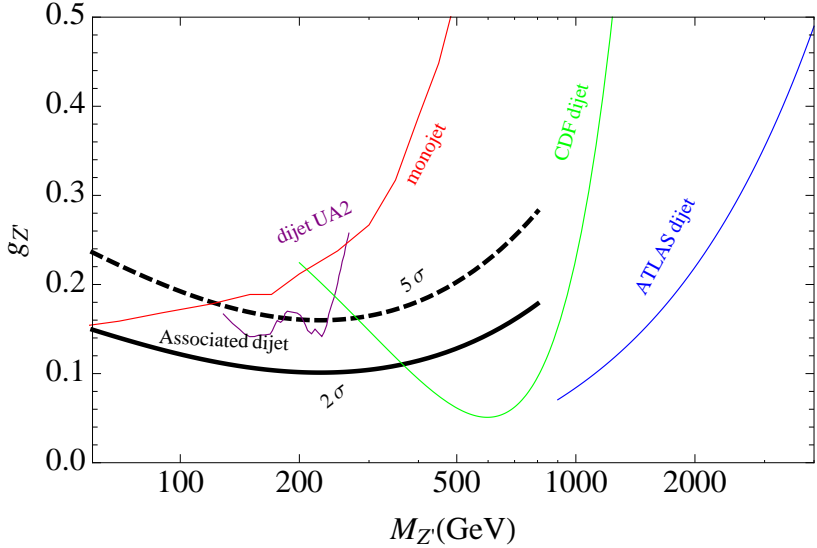}
\caption{Comparison between the combined constraint of different channels shown in Fig.~\ref{fig:Z4channels} and the constraints from dijet and monojet searches. We have assumed $r \equiv g_{D}/g_Z' =1$. The thick black curves shows the combined constraint as in Fig.~\ref{fig:Z4channels}. The red curve corresponds to the 95\% C.L.~ATLAS monojet upper limit. The green and blue curves correspond to 95\% C.L.~upper limits from CDF and ATLAS dijet searches, respectively. The purple curve corresponds to 90\% C.L.~upper limits from UA2 dijet search.}\label{fig:compareR}
\end{figure}

\begin{figure}
\centering
\includegraphics[height=3in]{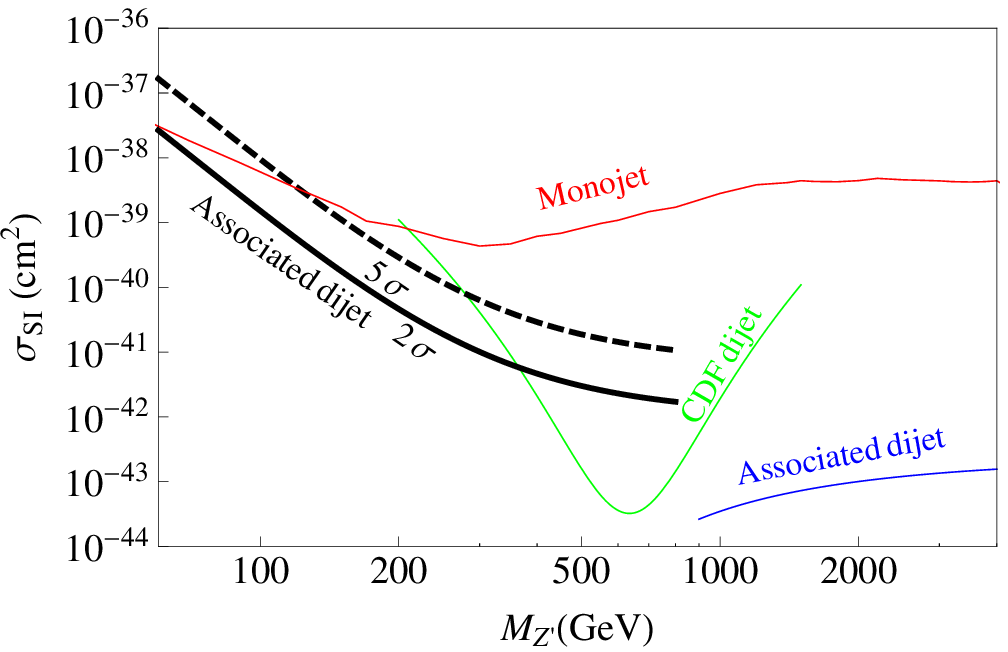}
\caption{Comparison of the $g_{Z'}=g_D$ reach of our $Z'$ association production (solid and dashed black curves for $2\sigma$ expected constraints and $5\sigma$ reach, respectively) with other experiments. The red curve is the constraint from ATLAS monojet search with 10 fb$^{-1}$ integrated luminosity at 8 TeV. The green curve is the bound from dijet resonance search by CDF with 1.13 fb$^{-1}$ integrated luminosity. The blue curve is the constraint from ATLAS dijet resonance search with 1 fb$^{-1}$ data set.}\label{fig:dijetplus}
\end{figure}

Fig.~\ref{fig:compareR} shows the comparison between the constraints from dijet search with associated products, monojet search and direct dijet searches. To make connection with dark matter detection, we will now consider $g_D \neq 0$  and introduce $r \equiv g_{D}/g_{Z'} $ to parameterize its size. As an illustration, we assume $r \equiv g_{D}/g_{Z'} = 1$ in Fig.~\ref{fig:compareR} . The red curve is the 95\% C.L. upper limit from ATLAS monojet searches~\cite{Atlas10fb}, the green and blue curves are the 95\% C.L. upper limit from the CDF~\cite{Aaltonen:2008dn} and ATLAS~\cite{Aad:2011fq} dijet resonance searches, and the purple curve shows the 90\%C.L. upper limit from UA2 dijet resonance search~\cite{Alitti:1993pn}, respectively. For dijet searches, colliders with smaller centre-of-mass energy give stronger constraints, since when $M_{Z'}$ is much smaller than the centre-of-mass energy of the collider, the constraint suffers from large QCD background due to the peak of the gluon parton distribution function at low $x$. For the comparison between the associated dijet constraint and the monojet constraint, a large amount of the background events can be removed with the help of the invariant mass window cut, so the associated dijet constraint can be stronger than the monojet constraint.  On the other hand, as discussed before, in the very light $Z'$ region ($M_Z'\lesssim80$ GeV), the two jets from the decay of $Z'$ are either highly boosted and cannot be distinguished from a single jet, or probably cut by the $p_{TJ}>50$ GeV threshold which is roughly half of the resonance mass, whereas the invisible decay of $Z'$ is only characterized by large missing transverse energy. Therefore, in this region the monojet constraints can be stronger than the associated dijet constraint.

Before the end of this section, we briefly mention some existing results in the $W^\pm\thinspace+$ dijet resonance channel.
Mainly motivated by checking the CDF $W^\pm\thinspace+$ dijet anomaly, the ATLAS \cite{Atlas102fb} and more recently CMS group \cite{:2012he} have performed searches in the same channel. In \cite{Atlas102fb} based on 5 fb$^{-1}$ integrated luminosity a leptophobic $Z'$ of 150 GeV and $g_{Z'}\simeq0.2$ \cite{Buckley:2011vc} is excluded. This is in broad agreement with our results shown in Fig.~\ref{fig:Z4channels}. Possible strategies to enhance the LHC reach in the $W^\pm\thinspace+$ dijet has also been studied in \cite{Eichten:2012hs}. However, the kinematics of this model in which the djiet resonance and $W$ are decay products of a heavier new resonance, is very different from the scenario considered in this paper.

\section{$Z'$ as a Portal between the SM and Dark Matter}
\label{sec:darkportal}

The $Z'$ can mediate interaction between dark matter and SM particles, forming the so called dark portal. In this case, the constraint on $g_{Z'}$ can be mapped onto the constraints on DM direct detection cross section. The direct detection cross section for a nucleon (proton or neutron) is
\begin{equation}\label{cxdirect}
\sigma_{SI}\simeq \frac{9g_{Z'}^2g_D^2M_N^2M_D^2}{\pi M_{Z'}^4(M_D+M_N)^2}\simeq7.7\times10^{-40}\left(\frac{g_{Z'}}{0.1}\right)^2\left(\frac{g_D}{0.1}\right)^2\left(\frac{100\mbox{GeV}}{M_{Z'}}\right)^4\thinspace\mbox{cm}^2,
\end{equation}
where $M_N$ is the mass of the nucleons, and $M_D=5$ GeV is assumed.

For $g_{Z'}=g_D$, the constraints on direct detections cross section are shown in Fig.~\ref{fig:dijetplus}.
The major improvement is in the region with $Z'$ lighter than limit from the CDF dijet pole search. The constraint can be as strong as a few $\times10^{-42}$ cm$^2$. Assuming $g_{Z'}=g_D$ and $M_{Z'}>80$ GeV, limits from associated production are also stronger than those from the ATLAS monojet search. The current bound assumes an integrated luminosity of 15 fb$^{-1}$, this constraint will becomes stronger if the assumed integrated luminosity increases.

Relaxing $g_{Z'}=g_D$ leads to interesting scaling behavior. The production rate of $Z'$ is proportional to $g_{Z'}^2$. The decay branching ratio of $Z'$ into dijet final states in the case of $g_{Z'}\neq g_D$ can be written as $g_{Z'}^2N_f N_C/(g_{Z'}^2N_f N_C + g_D^2)$. On the other hand, the decay branching ratio of $Z'$ into DM can be written as $g_D^2/(g_{Z'}^2N_f N_C + g_D^2)$. Therefore, for a general $r \equiv g_{D}/g_Z'$, the cross sections for monojet and dijet processes can be written as
\begin{eqnarray}\label{scaling}
\sigma_{\rm monojet} (r) &=& \sigma_{\rm monojet}^{(0)} \times \frac{N_C N_f +1}{N_C N_f +r^2} r^2  \propto g_{Z'} g_D\times r \frac{N_C N_f + 1}{N_C N_f + r^2}\ ; \nonumber \\
\sigma_{\rm dijet} (r) &=& \sigma_{\rm dijet}^{(0)} \times \frac{N_C N_f + 1}{N_C N_f +r^2} \propto g_{Z'} g_D \times \frac{1}{r} \frac{N_C N_f + 1}{N_C N_f +r^2} \ ,
\end{eqnarray}
where $\sigma_{\rm monojet}^{(0)}$ and $\sigma_{\rm dijet}^{(0)}$ are the cross sections for $r=1$. Therefore, as long as the $Z'$ is narrow-widthed and light enough so that it can be produced on shell, one can get the constraints on the coupling for a general value of $r$ by scaling the constraints shown in Fig.~\ref{fig:compareR} using Eq.~(\ref{scaling}).

\begin{figure}
\centering
\includegraphics[height=3in]{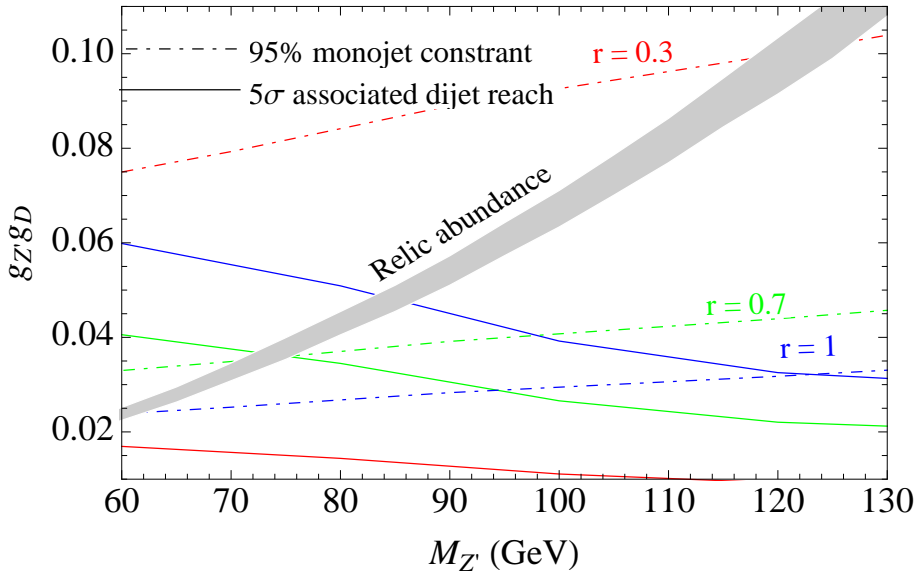}
\caption{Comparison of the 5$\sigma$ reach for the discovery of a 5 GeV WIMP by the $Z'$ associated production search (all solid curves) and monojet search (all dash-dotted curves), for different ratios of $r\equiv g_D/g_{Z'}$.  The region of parameter space above a certain curve either has been excluded or can be constrained by the corresponding search channel. All curves in red correspond to $g_D/g_{Z'} = 0.3$, all curves in green correspond to $g_D/g_{Z'} =0.7$, and all curves in blue correspond to $g_D/g_{Z'} =1$. In shaded region, the $\chi\bar{\chi}\leftrightarrow Z'$ process generates the correct relic abundance, namely requiring the cold dark matter density of the universe $\Omega_{\rm cdm} h^2 = 0.111\pm0.006$~\cite{pdg}. } \label{fig:withrelic}
\end{figure}

For fixed $g_{Z'}g_D$, the constraints of the production cross section from monojet and dijet searches have approximately opposite dependence on $r$. Therefore, the monojet method and our associated dijet method are complementary: the former works for large $r$ whereas the latter works for small $r$.
We can also take into account the theoretical consideration of DM thermal production, or the so called ``WIMP miracle''. If the $Z'$ is the only portal between the DM and the SM sectors, and the relic abundance of DM is determined by thermal freeze-out, the DM annihilation cross section to SM particles through the $Z'$ will be given in Eq.~(\ref{sigmav}), where we can see that $\langle\sigma v\rangle$ depends on the couplings only through the product $g_{Z'} g_D$ in the case that $M_D \ll M_{Z'}$. To get the observed relic abundance, $\langle\sigma v\rangle$ is fixed to be around $3\times10^{-26}$ cm$^3$s$^{-1}$; as a result, the product $g_{Z'}g_D$ becomes a function of $M_{Z'}$ for fixed $M_D$, which is shown as the grey band in Fig.~\ref{fig:withrelic}.

In the region that $M_{Z'}$ and $r$ are both small, the monojet constraint becomes weaker, as expected from Eq.~(\ref{scaling}). It cannot probe the shaded region in Fig.~\ref{fig:withrelic} where the correct relic abundance is generated. In the region $r<0.7$, from the crossing of the solid and dash dotted green curves in Fig.~\ref{fig:withrelic}, we can see that the correct relic abundance can be generated with parameters that satisfy the monojet constraint and can be reached by the associated dijet search with a significance larger than $5\sigma$ at the same time. In the region of $M_{Z'}<130$ GeV, from the red curves in Fig.~\ref{fig:withrelic}, we can see that if $r$ is smaller than about 0.3, the associated dijet resonance search can reach the full region of parameter space with correct relic abundance, with more than $5\sigma$ significance. Whereas in the region of $M_{Z'}>130$ GeV,  the associated dijet bump search cannot reach $5\sigma$ significance due to the constraint from UA2 dijet resonance search as shown in Fig.~\ref{fig:compareR}.

\section{Summary and discussions}

In this paper, we have suggested a new systematic way of finding/constraining the general leptophobic $Z'$ gauge boson. To overcome the large QCD background at the LHC, we considered the channels in which $Z'$s are produced in association with a hard jet, an energetic photon, or a $W^\pm/Z^0$.  We performed a detailed study of the potential of LHC search  in  these channels, and show that the $Z^\prime$ + jet, $Z'\thinspace+\thinspace\gamma$ and the $Z'\thinspace+\thinspace W^\pm$ channels are independently promising, and a combination can significantly improve the reach.

Assuming this $Z'$ also couples to dark matter, forming the so-called dark portal, this bound can be mapped onto a dark matter direct detection bound. We demonstrate that there is improvement in the low $Z'$ mass region ($M_{Z'}<400$ GeV) from the use of associated production. The collider constraint of DM direct detection cross section can be extended to a smaller hypothetical $Z'$ mass. For $g_{Z'}>g_D$, the $Z'$ associated production method holds an advantage over the monojet method. For example, if the $Z'$ mediates the only channel for DM to annihilate into SM particles, with small value of $r$, the monojet constraints become weak and the parameter space for generating the correct relic abundance with small 5 GeV $M_D$ and weak scale $Z'$ is still allowed, as shown in Fig.~\ref{fig:withrelic}. At the same time, large region of this parameter space can be covered by the associated dijet bump search discussed in this paper. In particular, we can discover such a $Z'$ in this search channel with a significance larger than $5 \sigma$,  if $g_D/g_{Z'}$ is smaller than about 0.7.

\section{Acknowledgments}\label{Acknowledgments}

We would like to acknowledge useful information from Antonio Boveia about the triggering of the ATLAS experiment. R.H. wish to thank Peter Skands and Xiaohui Liu for useful discussions. H.A. is supported by in part by the Government of Canada through NSERC and by the Province of Ontario through MEDT. L.T.W. is supported by the NSF under grant PHY-0756966 and the DOE Early Career Award under grant DE-SC0003930.

\appendix

\section{Variations on jet energy thresholds}
\label{sec:jet_trig}

\begin{table}[h!]
\centering
\begin{tabular}{|c|p{2.2cm}|p{2.2cm}|p{2.2cm}|p{2.2cm}|p{2.2cm}|}
\hline
Threshold & 350-70-70 & 300-70-70 & 400-70-70 & 350-60-60 & 350-80-80 \\
Background & 370~pb & 767~pb & 200~pb & 410~pb & 333~pb \\
$S/\sqrt{B}$ & 15.0 & 19.4 & 8.9 & 15.2 & 14.8 \\
\hline
\end{tabular}
\caption{\label{TrigVar_1} Comparison of small variation of the first set of jet energy thresholds for the 3-jet signal. For the leading order event rate we also impose the $|\Delta\eta_{23}|<1.7$ cut, but no invariant mass window choice. In $S/\sqrt{B}$ we are assuming $g_L=g_R=0.35$, $M_{Z'}=150$ GeV and 15 fb$^{-1}$ integrated luminosity.}
\end{table}
\medskip
\begin{table}[h!]
\centering
\begin{tabular}{|c|p{2.2cm}|p{2.2cm}|p{2.2cm}|p{2.2cm}|p{2.2cm}|}
\hline
Trigger level & 160-160-160 & 150-150-150 & 140-140-140 & 170-170-170 & 180-180-180 \\
Background & 286~pb & 406~pb & 579~pb & 205~pb & 148~pb \\
$S/\sqrt{B}$ & 10.8 & 11.2 & 12.6 & 8.9 & 8.0 \\
\hline
\end{tabular}
\caption{\label{TrigVar_2}  Comparison of small variation of the second set of jet energy thresholds for the 3-jet signal. For the leading order event rate we also impose the $|\Delta\eta_{12}|<1.7$ cut, but no invariant mass window choice. In $S/\sqrt{B}$ we are assuming $g_L=g_R=0.35$, $M_{Z'}=450$ GeV and 15 fb$^{-1}$ integrated luminosity.}
\end{table}
We compare different choices of jet energy thresholds, and the results are tabulated here.

\section{Low $M_{Z'}$ detection thresholds for different channels}
\label{appendixB}

\begin{figure}
\centering
\includegraphics[height=3in]{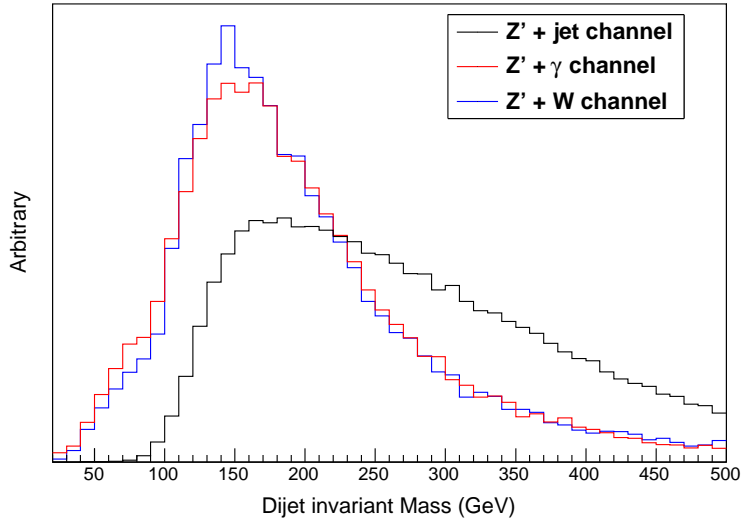}
\caption{Background distributions of dijet invariant mass in $Z'$+jet, $Z'+\gamma$ and $Z'+W$ channels. For $Z'+$jet channel, it is the invariant mass of the second and third hardest jets shown in the plot. } \label{fig:distribution}
\end{figure}

We start with considering an event with three jets  in the final state, labelled by $J_1$, $J_2$ and $J_3$,  with transverse momenta $\vec p_{T1}$, $\vec p_{T2}$ and $\vec p_{T3}$, respectively. Here, we require that $p_{T1}\geq p_{T2}  \geq  p_{T3}$. Then, the invariant mass of $J_2$ and $J_3$ can be written as
\begin{equation}
M_{23}^2 = 2(p_{T2}^2+p_{z2}^2)^{1/2} (p_{T3}^2+p_{z3}^2)^{1/2} - 2 \vec p_{T2} \cdot \vec p_{T3} - 2 p_{2z}p_{3z} \ ,
\end{equation}
where we neglect the invariant mass of a single jet. $p_{2z}$ and $p_{3z}$ are the longitudinal momenta of $J_2$ and $J_3$, respectively. We can always work in a frame where $p_{z3} = 0$ such that the above expression can be simplified as
\begin{eqnarray}
M_{23}^2 &=& 2(p_{T2}^2+p_{z2}^2)^{1/2} p_{T3}- 2 \vec p_{T2} \cdot \vec p_{T3} \nonumber \\
                  &=& 2 p_{T2} p_{T3}  \left[ \left( 1 + \frac{p_{z2}^2}{p_{T2}^2}\right)^{1/2} - 1\right] + 2p_{2T}p_{3T}(1-\cos\phi) \ ,
\end{eqnarray}
where
$\phi$ is the angle between $\vec p_{2T}$ and $\vec p_{3T}$. It is easy to see that $2 p_{T2}p_{T3} > 2 p_{T3}^{\rm min} (p_{T1}^{\rm min} - p_{T3}^{\rm min})\approx (200 {\rm GeV})^2$, where $p_{T1}^{\rm min}$, $p_{T2}^{\rm min}$ and $p_{T3}^{\rm min}$ are the cuts we impose on $p_{T1}$, $p_{T2}$ and $p_{T3}$. In practice, for the region $M_{Z'} < 350$ GeV, we choose $p_{T1}^{\rm min} = 350$ GeV, $p_{T2}^{\rm min}= p_{T3}^{\rm min}= 70$ GeV. Therefore, in the region $M_{23}^2\ll (200 {\rm GeV})^2$ we have
\begin{eqnarray}
M_{23}^2 &\approx& 2p_{T2} p_{T3} \left( \frac{p_{z2}^2}{2p_{T2}^2} + \frac{1}{2} \phi^2\right) \nonumber \\
                  &\approx& p_{T2} p_{T3} (y^2 + \phi^2) \ ,
\end{eqnarray}
where $y$ is the rapidity of $J_3$. In the anti-k$_T$ algorithm, the distance between $J_2$ and $J_3$ is defined as
\begin{equation}
d_{23} = \min\left({p^{-2}_{T2}},{p^{-2}_{T3}}\right) \frac{y^2 + \phi^2}{R^2} \approx \frac{M_{23}^2}{p_{T2}^3 p_{T3} R^2} \ .
\end{equation}
If $d_{23} < 1/p_{T2}^2$, $J_2$ and $J_3$ will be identified as a single jet. Therefore, it requires $d_{23} > 1/p_{T2}^2$ to identify $J_2$ and $J_3$ as two separate jets, which means
\begin{equation}
M_{23} > R\sqrt{p_{2T}p_{3T}} \geq R\sqrt{p_{T3}^{\rm min} (p^{\rm min}_{T1} - p^{\rm min}_{T3})} \approx 80 {~\rm GeV} \ ,
\end{equation}
where $R = 0.6$ has been used. Therefore, the cuts we imposed on the three-jet final state configuration remove both the background and the signal in small $M_{23}$ region as shown in Fig.~\ref{fig:distribution}. We should note that this argument is based on parton level analysis. Taking the parton shower and realistic detector effects in to account, a small amount of events from both background and signal in the region $M_{23} < R\sqrt{p_{T3}^{\rm min} (p^{\rm min}_{T1} - p^{\rm min}_{T3})}$ may pass the cuts and leak into the signal region. However, the statistics in the small $M_{23}$ region remains very poor, and so does the constraint on the coupling.

For the $Z'+\gamma$ channel, as shown in Fig.~\ref{fig:distribution}, since we don't require a very large transverse momentum for the photon, the limit of the invariant mass of the two jets can simply be written as
\begin{equation}
M_{12} \geq 2 p^{\rm min}_{TJ} R / 2 \approx 30 ~{\rm GeV} \ ,
\end{equation}
where $p^{\rm min}_{TJ} = 50$ GeV is used. Therefore, the constraint from this channel dominates over the $Z'+$jet channel in the small $M_{Z'}$ region.

For the $Z'+W$ channel, since we only consider the leptonic channel, $W$ boson does not have to be produced with a large transverse boost to pass the selection cuts. Therefore, $Z'$ don't have to be very boosted, and there will be no limitation on $M_{12}$ in the small $M_{Z'}$ region.
For the $Z'+Z^0$ channel, however, the dominant decay channel of $Z^0$ we are considering is to a pair of neutrinos and the $\not\!\!p_T > 60$ is required, which indicates that the $p_T$ of $Z^0$ boson must be larger than 60 GeV. For fixed dijet invariant mass $M_{12}$, in the $Z'+W$ channel, ${\hat s}^{1/2}_{\rm min} = M_{W}+ M_{12}$, where ${\hat s}^{1/2}_{\rm min}$ is the minimal parton level centre-of-mass energy to make this process possible. Whereas in the $Z'+Z^0$ channel, ${\hat s}^{1/2}_{\rm min} = \sqrt{M_Z^2 + {p_{TZ}^{\rm min}}^2} + \sqrt{M_{Z'}^2 + {p_{TZ}^{\rm min}}^2}$. Therefore, for both signal and background, in the small $M_{Z'}$ region, ${\hat s}^{1/2}_{\rm min}$ decreases faster in the $Z'+W$ channel than in the $Z'+Z^0$ channel with the decreasing of $M_{Z'}$. As a result, the parton luminosity increases faster in the $Z'+W$ channel than in the $Z'+Z^0$ channel with the decreasing of $M_{Z'}$. Thus, the upper limit on $g_{Z'}$ from the $Z+Z'$ channel increases more rapidly in the small $M_{Z'}$ region with the decreasing of $M_{Z'}$ as shown in Fig.~\ref{fig:Z4channels}.


\begin{thebibliography}{99}

\bibitem{Goodman:1984dc}
  M.~W.~Goodman and E.~Witten,
  ``Detectability of Certain Dark Matter Candidates,''  Phys.\ Rev.\ D {\bf 31}, 3059 (1985).  

\bibitem{Aalseth:2011wp}
  C.~E.~Aalseth, P.~S.~Barbeau, J.~Colaresi, J.~I.~Collar, J.~Diaz Leon, J.~E.~Fast, N.~Fields and T.~W.~Hossbach {\it et al.},
  ``Search for an Annual Modulation in a P-type Point Contact Germanium Dark Matter Detector,''  Phys.\ Rev.\ Lett.\  {\bf 107}, 141301 (2011)  [arXiv:1106.0650 [astro-ph.CO]].  

\bibitem{Akerib:2010pv}
  D.~S.~Akerib {\it et al.}  [CDMS Collaboration],
  ``A low-threshold analysis of CDMS shallow-site data,''  Phys.\ Rev.\ D {\bf 82}, 122004 (2010)  [arXiv:1010.4290 [astro-ph.CO]].  

\bibitem{Ahmed:2010wy}
  Z.~Ahmed {\it et al.}  [CDMS-II Collaboration],
  ``Results from a Low-Energy Analysis of the CDMS II Germanium Data,''  Phys.\ Rev.\ Lett.\  {\bf 106}, 131302 (2011)  [arXiv:1011.2482 [astro-ph.CO]].  

\bibitem{Aprile:2010um}
  E.~Aprile {\it et al.}  [XENON100 Collaboration],
  ``First Dark Matter Results from the XENON100 Experiment,''  Phys.\ Rev.\ Lett.\  {\bf 105}, 131302 (2010)  [arXiv:1005.0380 [astro-ph.CO]].  

\bibitem{Angloher:2011uu}
  G.~Angloher, M.~Bauer, I.~Bavykina, A.~Bento, C.~Bucci, C.~Ciemniak, G.~Deuter and F.~von Feilitzsch {\it et al.},
  ``Results from 730 kg days of the CRESST-II Dark Matter Search,''  Eur.\ Phys.\ J.\ C {\bf 72}, 1971 (2012)  [arXiv:1109.0702 [astro-ph.CO]].  

\bibitem{Bernabei:2008yi}
  R.~Bernabei {\it et al.}  [DAMA Collaboration],
  ``First results from DAMA/LIBRA and the combined results with DAMA/NaI,''  Eur.\ Phys.\ J.\ C {\bf 56}, 333 (2008)  [arXiv:0804.2741 [astro-ph]].  

\bibitem{Aprile:2011hi}
  E.~Aprile {\it et al.}  [XENON100 Collaboration],
  ``Dark Matter Results from 100 Live Days of XENON100 Data,''  Phys.\ Rev.\ Lett.\  {\bf 107}, 131302 (2011)  [arXiv:1104.2549 [astro-ph.CO]].  

\bibitem{Felizardo:2010mi}
  M.~Felizardo, T.~Morlat, A.~C.~Fernandes, T.~A.~Girard, J.~G.~Marques, A.~R.~Ramos, M.~Auguste and D.~Boyer {\it et al.},
  ``First Results of the Phase II SIMPLE Dark Matter Search,''  Phys.\ Rev.\ Lett.\  {\bf 105}, 211301 (2010)  [arXiv:1003.2987 [astro-ph.CO]].  

\bibitem{Aprile:2012nq}
  [XENON100 Collaboration],
  ``Dark Matter Results from 225 Live Days of XENON100 Data,''  arXiv:1207.5988 [astro-ph.CO].  

\bibitem{Nussinov:1985xr}
  S.~Nussinov,
  Phys.\ Lett.\ B {\bf 165}, 55 (1985).

\bibitem{Kaplan:2009ag}
  D.~E.~Kaplan, M.~A.~Luty and K.~M.~Zurek,
  ``Asymmetric Dark Matter,''  Phys.\ Rev.\ D {\bf 79}, 115016 (2009)  [arXiv:0901.4117 [hep-ph]].  

\bibitem{Farrar:2005zd}
  G.~R.~Farrar and G.~Zaharijas,
  ``Dark matter and the baryon asymmetry,''  Phys.\ Rev.\ Lett.\  {\bf 96}, 041302 (2006)  [hep-ph/0510079].  

\bibitem{Kitano:2004sv}
  R.~Kitano and I.~Low,
  ``Dark matter from baryon asymmetry,''  Phys.\ Rev.\ D {\bf 71}, 023510 (2005)  [hep-ph/0411133].  

\bibitem{Agashe:2004ci}
  K.~Agashe and G.~Servant,
  ``Warped unification, proton stability and dark matter,''  Phys.\ Rev.\ Lett.\  {\bf 93}, 231805 (2004)  [hep-ph/0403143].  

\bibitem{An:2009vq}
  H.~An, S.~-L.~Chen, R.~N.~Mohapatra and Y.~Zhang,
  ``Leptogenesis as a Common Origin for Matter and Dark Matter,''  JHEP {\bf 1003}, 124 (2010)  [arXiv:0911.4463 [hep-ph]].  

\bibitem{Shelton:2010ta}
  J.~Shelton and K.~M.~Zurek,
  ``Darkogenesis: A baryon asymmetry from the dark matter sector,''  Phys.\ Rev.\ D {\bf 82}, 123512 (2010)  [arXiv:1008.1997 [hep-ph]].  

\bibitem{Davoudiasl:2010am}
  H.~Davoudiasl, D.~E.~Morrissey, K.~Sigurdson and S.~Tulin,
  ``Hylogenesis: A Unified Origin for Baryonic Visible Matter and Antibaryonic Dark Matter,''  Phys.\ Rev.\ Lett.\  {\bf 105}, 211304 (2010)  [arXiv:1008.2399 [hep-ph]].  

\bibitem{Bell:2011tn}
  N.~F.~Bell, K.~Petraki, I.~M.~Shoemaker and R.~R.~Volkas,
  ``Pangenesis in a Baryon-Symmetric Universe: Dark and Visible Matter via the Affleck-Dine Mechanism,''  Phys.\ Rev.\ D {\bf 84}, 123505 (2011)  [arXiv:1105.3730 [hep-ph]].  

\bibitem{Gu:2010ft}
  P.~-H.~Gu, M.~Lindner, U.~Sarkar and X.~Zhang,
  ``WIMP Dark Matter and Baryogenesis,''  Phys.\ Rev.\ D {\bf 83}, 055008 (2011)  [arXiv:1009.2690 [hep-ph]].  

\bibitem{Blennow:2010qp}
  M.~Blennow, B.~Dasgupta, E.~Fernandez-Martinez and N.~Rius,
  ``Aidnogenesis via Leptogenesis and Dark Sphalerons,''  JHEP {\bf 1103}, 014 (2011)  [arXiv:1009.3159 [hep-ph]].  

\bibitem{Dutta:2010va}
  B.~Dutta and J.~Kumar,
  ``Asymmetric Dark Matter from Hidden Sector Baryogenesis,''  Phys.\ Lett.\ B {\bf 699}, 364 (2011)  [arXiv:1012.1341 [hep-ph]].  

\bibitem{Kang:2011wb}
  Z.~Kang, J.~Li, T.~Li, T.~Liu and J.~Yang,
  ``Asymmetric Sneutrino Dark Matter in the NMSSM with Minimal Inverse Seesaw,''  arXiv:1102.5644 [hep-ph].  

\bibitem{Cheung:2011if}
  C.~Cheung and K.~M.~Zurek,
  ``Affleck-Dine Cogenesis,''  Phys.\ Rev.\ D {\bf 84}, 035007 (2011)  [arXiv:1105.4612 [hep-ph]].  

\bibitem{MarchRussell:2011fi}
  J.~March-Russell and M.~McCullough,
  ``Asymmetric Dark Matter via Spontaneous Co-Genesis,''  JCAP {\bf 1203}, 019 (2012)  [arXiv:1106.4319 [hep-ph]].  

\bibitem{Frandsen:2011kt}
  M.~T.~Frandsen, S.~Sarkar and K.~Schmidt-Hoberg,
  ``Light asymmetric dark matter from new strong dynamics,''  Phys.\ Rev.\ D {\bf 84}, 051703 (2011)  [arXiv:1103.4350 [hep-ph]].  

\bibitem{Ibe:2011hq}
  M.~Ibe, S.~Matsumoto and T.~T.~Yanagida,
  ``The GeV-scale dark matter with B-L asymmetry,''  Phys.\ Lett.\ B {\bf 708}, 112 (2012)  [arXiv:1110.5452 [hep-ph]].  

\bibitem{Kamada:2012ht}
  K.~Kamada and M.~Yamaguchi,
  ``Asymmetric Dark Matter from Spontaneous Cogenesis in the Supersymmetric Standard Model,''  Phys.\ Rev.\ D {\bf 85}, 103530 (2012)  [arXiv:1201.2636 [hep-ph]].  

\bibitem{Feng:2012jn}
  W.~-Z.~Feng, P.~Nath and G.~Peim,
  ``Cosmic Coincidence and Asymmetric Dark Matter in a Stueckelberg Extension,''  Phys.\ Rev.\ D {\bf 85}, 115016 (2012)  [arXiv:1204.5752 [hep-ph]].  

\bibitem{Goodman:2010yf}
  J.~Goodman, M.~Ibe, A.~Rajaraman, W.~Shepherd, T.~M.~P.~Tait and H.~-B.~Yu,
  ``Constraints on Light Majorana dark Matter from Colliders,''  Phys.\ Lett.\ B {\bf 695}, 185 (2011)  [arXiv:1005.1286 [hep-ph]].  

\bibitem{Bai:2010hh}
  Y.~Bai, P.~J.~Fox and R.~Harnik,
  ``The Tevatron at the Frontier of Dark Matter Direct Detection,''  JHEP {\bf 1012}, 048 (2010)  [arXiv:1005.3797 [hep-ph]].  

\bibitem{Goodman:2010ku}
  J.~Goodman, M.~Ibe, A.~Rajaraman, W.~Shepherd, T.~M.~P.~Tait and H.~-B.~Yu,
  ``Constraints on Dark Matter from Colliders,''  Phys.\ Rev.\ D {\bf 82}, 116010 (2010)  [arXiv:1008.1783 [hep-ph]].  

\bibitem{Fortin:2011hv}
  J.~-F.~Fortin and T.~M.~P.~Tait,
  ``Collider Constraints on Dipole-Interacting Dark Matter,''  Phys.\ Rev.\ D {\bf 85}, 063506 (2012)  [arXiv:1103.3289 [hep-ph]].  

\bibitem{Rajaraman:2011wf}
  A.~Rajaraman, W.~Shepherd, T.~M.~P.~Tait and A.~M.~Wijangco,
  ``LHC Bounds on Interactions of Dark Matter,''  Phys.\ Rev.\ D {\bf 84}, 095013 (2011)  [arXiv:1108.1196 [hep-ph]].  

\bibitem{Shoemaker:2011vi}
  I.~M.~Shoemaker and L.~Vecchi,
  ``Unitarity and Monojet Bounds on Models for DAMA, CoGeNT, and CRESST-II,''  arXiv:1112.5457 [hep-ph].  

\bibitem{Graesser:2011vj}
  M.~L.~Graesser, I.~M.~Shoemaker and L.~Vecchi,
  ``A Dark Force for Baryons,''  arXiv:1107.2666 [hep-ph].  

\bibitem{Friedland:2011za}
  A.~Friedland, M.~L.~Graesser, I.~M.~Shoemaker and L.~Vecchi,
  ``Probing Nonstandard Standard Model Backgrounds with LHC Monojets,''  Phys.\ Lett.\ B {\bf 714}, 267 (2012)  [arXiv:1111.5331 [hep-ph]].  

\bibitem{An:2012va}
  H.~An, X.~Ji and L.~-T.~Wang,
  ``Light dark matter and $Z'$ dark force at colliders,''
  arXiv:1202.2894 [hep-ph].  

\bibitem{Alwall:2011uj}
  J.~Alwall, M.~Herquet, F.~Maltoni, O.~Mattelaer and T.~Stelzer,
  ``MadGraph 5 : Going Beyond,''  JHEP {\bf 1106}, 128 (2011)  [arXiv:1106.0522 [hep-ph]].  

\bibitem{Meade:2007js}
  P.~Meade and M.~Reece,
  ``BRIDGE: Branching ratio inquiry / decay generated events,''  hep-ph/0703031.  

\bibitem{Cacciari:2011ma}
  M.~Cacciari, G.~P.~Salam and G.~Soyez,
  ``FastJet user manual,''  Eur.\ Phys.\ J.\ C {\bf 72}, 1896 (2012)  [arXiv:1111.6097 [hep-ph]].  

\bibitem{Ball:2007zza}
  G.~L.~Bayatian {\it et al.}  [CMS Collaboration],
  ``CMS technical design report, volume II: Physics performance,''  J.\ Phys.\ G G {\bf 34}, 995 (2007).  

\bibitem{Campbell:MCFM}
  J.~Campbell, K.~Ellis and C.~Williams,
  http://mcfm.fnal.gov/.  

\bibitem{Bern:2011ep}
  Z.~Bern, G.~Diana, L.~J.~Dixon, F.~Febres Cordero, S.~Hoeche, D.~A.~Kosower, H.~Ita and D.~Maitre {\it et al.},
  ``Four-Jet Production at the Large Hadron Collider at Next-to-Leading Order in QCD,''  arXiv:1112.3940 [hep-ph].  

\bibitem{Bern:2011pa}
  Z.~Bern, G.~Diana, L.~J.~Dixon, F.~Febres Cordero, S.~Hoche, H.~Ita, D.~A.~Kosower and D.~Maitre {\it et al.},
  ``Driving Missing Data at Next-to-Leading Order,''  Phys.\ Rev.\ D {\bf 84}, 114002 (2011)  [arXiv:1106.1423 [hep-ph]].  


\bibitem{Altheimer:2012mn}
  A.~Altheimer, S.~Arora, L.~Asquith, G.~Brooijmans, J.~Butterworth, M.~Campanelli, B.~Chapleau and A.~E.~Cholakian {\it et al.},
  J.\ Phys.\ G {\bf 39}, 063001 (2012)
  [arXiv:1201.0008 [hep-ph]].


\bibitem{Atlas10fb}
  [ATLAS Collaboration],
  ATLAS-CONF-2012-147



\bibitem{Aaltonen:2008dn}
  T.~Aaltonen {\it et al.}  [CDF Collaboration],
  Phys.\ Rev.\ D {\bf 79}, 112002 (2009)
  [arXiv:0812.4036 [hep-ex]].

\bibitem{Aad:2011fq}
  G.~Aad {\it et al.}  [ATLAS Collaboration],
  Phys.\ Lett.\ B {\bf 708}, 37 (2012)
  [arXiv:1108.6311 [hep-ex]].

\bibitem{Alitti:1993pn}
  J.~Alitti {\it et al.}  [UA2 Collaboration],
  Nucl.\ Phys.\ B {\bf 400}, 3 (1993).

\bibitem{Atlas102fb}
  [ATLAS Collaboration],
  ATLAS-CONF-2011-097

\bibitem{:2012he}
  S.~Chatrchyan {\it et al.}  [CMS Collaboration],
  ``Study of the dijet mass spectrum in $pp \to W +$ jets events at $\sqrt{s}=7$ TeV,''  arXiv:1208.3477 [hep-ex].  

\bibitem{Buckley:2011vc}
  M.~R.~Buckley, D.~Hooper, J.~Kopp and E.~Neil,
  ``Light Z' Bosons at the Tevatron,''  Phys.\ Rev.\ D {\bf 83}, 115013 (2011)  [arXiv:1103.6035 [hep-ph]].  

\bibitem{Eichten:2012hs}
  E.~Eichten, K.~Lane, A.~Martin and E.~Pilon,
  Phys.\ Rev.\ D {\bf 86}, 074015 (2012)
  [arXiv:1206.0186 [hep-ph]].

\bibitem{pdg}
 ÊJ.~Beringer {\it et al.} Ê[Particle Data Group Collaboration],
 Ê``Review of Particle Physics (RPP),'' ÊPhys.\ Rev.\ D {\bf 86}, 010001 (2012). Ê

\end{thebibliography}
\end{document}